\PassOptionsToPackage{unicode}{hyperref}
\PassOptionsToPackage{hyphens}{url}
\PassOptionsToPackage{dvipsnames,svgnames,x11names}{xcolor}
\documentclass[
]{article}
\usepackage{xcolor}
\usepackage[top=20mm,left=20mm,right=20mm,bottom=20mm]{geometry}
\usepackage{amsmath,amssymb}
\usepackage{iftex}
\ifPDFTeX
  \usepackage[T1]{fontenc}
  \usepackage[utf8]{inputenc}
  \usepackage{textcomp} 
\else 
  \usepackage{unicode-math} 
  \defaultfontfeatures{Scale=MatchLowercase}
  \defaultfontfeatures[\rmfamily]{Ligatures=TeX,Scale=1}
\fi
\usepackage{lmodern}
\ifPDFTeX\else
\fi
\IfFileExists{upquote.sty}{\usepackage{upquote}}{}
\IfFileExists{microtype.sty}{
  \usepackage[]{microtype}
  \UseMicrotypeSet[protrusion]{basicmath} 
}{}
\makeatletter
\@ifundefined{KOMAClassName}{
  \IfFileExists{parskip.sty}{%
    \usepackage{parskip}
  }{
    \setlength{\parindent}{0pt}
    \setlength{\parskip}{6pt plus 2pt minus 1pt}}
}{
  \KOMAoptions{parskip=half}}
\makeatother
\makeatletter
\ifx\paragraph\undefined\else
  \let\oldparagraph\paragraph
  \renewcommand{\paragraph}{
    \@ifstar
      \xxxParagraphStar
      \xxxParagraphNoStar
  }
  \newcommand{\xxxParagraphStar}[1]{\oldparagraph*{#1}\mbox{}}
  \newcommand{\xxxParagraphNoStar}[1]{\oldparagraph{#1}\mbox{}}
\fi
\ifx\subparagraph\undefined\else
  \let\oldsubparagraph\subparagraph
  \renewcommand{\subparagraph}{
    \@ifstar
      \xxxSubParagraphStar
      \xxxSubParagraphNoStar
  }
  \newcommand{\xxxSubParagraphStar}[1]{\oldsubparagraph*{#1}\mbox{}}
  \newcommand{\xxxSubParagraphNoStar}[1]{\oldsubparagraph{#1}\mbox{}}
\fi
\makeatother

\usepackage{longtable,booktabs,array}
\usepackage{calc} 
\usepackage{etoolbox}
\makeatletter
\patchcmd\longtable{\par}{\if@noskipsec\mbox{}\fi\par}{}{}
\makeatother
\IfFileExists{footnotehyper.sty}{\usepackage{footnotehyper}}{\usepackage{footnote}}
\makesavenoteenv{longtable}
\usepackage{graphicx}
\makeatletter
\newsavebox\pandoc@box
\newcommand*\pandocbounded[1]{
  \sbox\pandoc@box{#1}%
  \Gscale@div\@tempa{\textheight}{\dimexpr\ht\pandoc@box+\dp\pandoc@box\relax}%
  \Gscale@div\@tempb{\linewidth}{\wd\pandoc@box}%
  \ifdim\@tempb\p@<\@tempa\p@\let\@tempa\@tempb\fi
  \ifdim\@tempa\p@<\p@\scalebox{\@tempa}{\usebox\pandoc@box}%
  \else\usebox{\pandoc@box}%
  \fi%
}
\def\fps@figure{htbp}
\makeatother

\providecommand{\tightlist}{%
  \setlength{\itemsep}{0pt}\setlength{\parskip}{0pt}}

\usepackage[square,numbers,sort,compress]{natbib}
\usepackage{xcolor}
\usepackage{braket}

\usepackage{url}

\usepackage[bookmarks=false]{hyperref}
\hypersetup{
  colorlinks=true,
  linkcolor=blue,
  filecolor=magenta,
  urlcolor=cyan,
}

\usepackage{float}

\usepackage{orcidlink}
\definecolor{mypink}{RGB}{219, 48, 122}
\makeatletter
\@ifpackageloaded{caption}{}{\usepackage{caption}}
\AtBeginDocument{%
\ifdefined\contentsname
  \renewcommand*\contentsname{Table of contents}
\else
  \newcommand\contentsname{Table of contents}
\fi
\ifdefined\listfigurename
  \renewcommand*\listfigurename{List of Figures}
\else
  \newcommand\listfigurename{List of Figures}
\fi
\ifdefined\listtablename
  \renewcommand*\listtablename{List of Tables}
\else
  \newcommand\listtablename{List of Tables}
\fi
\ifdefined\figurename
  \renewcommand*\figurename{Figure}
\else
  \newcommand\figurename{Figure}
\fi
\ifdefined\tablename
  \renewcommand*\tablename{Table}
\else
  \newcommand\tablename{Table}
\fi
}
\@ifpackageloaded{float}{}{\usepackage{float}}
\floatstyle{ruled}
\@ifundefined{c@chapter}{\newfloat{codelisting}{h}{lop}}{\newfloat{codelisting}{h}{lop}[chapter]}
\floatname{codelisting}{Listing}

\makeatother
\makeatletter
\@ifpackageloaded{caption}{}{\usepackage{caption}}
\@ifpackageloaded{subcaption}{}{\usepackage{subcaption}}
\makeatother
\usepackage{bookmark}
\IfFileExists{xurl.sty}{\usepackage{xurl}}{} 
\hypersetup{
  pdftitle={Awesome Quantum Computing Experiments: Benchmarking Experimental Progress Towards Fault-Tolerant Quantum Computation},
  pdfauthor={François-Marie Le Régent},
  colorlinks=true,
  linkcolor={blue},
  filecolor={Maroon},
  citecolor={Blue},
  urlcolor={blue},
  pdfcreator={LaTeX via pandoc}}

\title{Awesome Quantum Computing Experiments: Benchmarking Experimental
Progress Towards Fault-Tolerant Quantum Computation}
\author{
François-Marie Le
Régent\\Pasqal\\\href{mailto:francois-marie.le-regent@pasqal.com}{francois-marie.le-regent@pasqal.com}}
\date{}
\begin{document}
\maketitle
\begin{abstract}
Achieving fault-tolerant quantum computation (FTQC) demands simultaneous
progress in physical qubit performance and quantum error correction
(QEC). This work reviews and benchmarks experimental advancements
towards FTQC across leading platforms, including trapped ions,
superconducting circuits, neutral atoms, NV centers, and semiconductors.
We analyze key physical metrics like coherence times, entanglement
error, and system size (qubit count), fitting observed exponential
trends to characterize multi-order-of-magnitude improvements over the
past two decades. At the logical
level, we survey the implementation landscape of QEC codes, tracking
realized parameters \([[n, k, d]]\) and complexity from early
demonstrations to recent surface and color code experiments.
Synthesizing these physical and logical benchmarks reveals substantial
progress enabled by underlying hardware improvements, while also
outlining persistent challenges towards scalable FTQC. The experimental
databases and analysis code underpinning this review are publicly
available at
\url{https://github.com/francois-marie/awesome-quantum-computing-experiments}.
\end{abstract}

\renewcommand*\contentsname{Table of contents}
{
\hypersetup{linkcolor=}
\setcounter{tocdepth}{3}
\tableofcontents
}

\section{Introduction}\label{introduction}

The realization of algorithms offering quantum advantage
\cite{lloydUniversalQuantumSimulators1996,  shorPolynomialTimeAlgorithmsPrime1997,  farhiQuantumAdiabaticEvolution2001, aspuruguzikSimulatedQuantumComputation2005}
demands quantum processors operating not just with a large number of
qubits
\cite{gidneyHowFactor20482021a, omanakuttanThresholdFaulttolerantQuantum2025, gidneyHowFactor20482025, zhouResourceAnalysisLowOverhead2025},
but with qubits exhibiting high fidelity, often called ``algorithmic
qubits'' (qubits meeting the fidelity requirements for useful
algorithms)
\cite{kivlichanImprovedFaultTolerantQuantum2020, campbellEarlyFaulttolerantSimulations2022}.
Early physical qubit implementations
\cite{coryExperimentalQuantumError1998,turchetteDeterministicEntanglementTwo1998,nakamuraCoherentControlMacroscopic1999a},
while demonstrating fundamental quantum phenomena, suffered from
significant error rates, far exceeding the requirements for complex
computations
\cite{shorFaulttolerantQuantumComputation1996, kimFaulttolerantResourceEstimate2022}.
The primary strategy to bridge this gap lies in quantum error correction
(QEC)
\cite{shorFaulttolerantQuantumComputation1996,steaneMultipleParticleInterference1996,dennisTopologicalQuantumMemory2002},
where quantum information is redundantly encoded across multiple
physical qubits to protect it from local errors, effectively creating
robust logical qubits.

Progress towards fault-tolerant quantum computation
\cite{shorSchemeReducingDecoherence1995,steaneSimpleQuantumErrorcorrecting1996,gottesmanStabilizerCodesQuantum1997a,knillTheoryQuantumErrorcorrecting1997,preskillFaulttolerantQuantumComputation1997,
knillResilientQuantumComputation1998b} is thus intrinsically
multi-faceted, requiring simultaneous improvements across all layers of
the hardware stack. Key areas include extending the fundamental
lifetimes (\(T_1\)) and coherence times (\(T_2\)) of physical qubits,
minimizing errors in single- and multi-qubit gate operations,
dramatically increasing the number of addressable and interconnected
physical qubits, and successfully implementing and scaling QEC codes of
increasing complexity and distance.

While individual metrics offer valuable insights into specific aspects
of a given quantum computing platform's maturity
\cite{kjaergaardSuperconductingQubitsCurrent2020}, a holistic view
combining physical-level benchmarks with logical-level QEC
implementation progress \cite{albertQuantumRealizationsError2025} is
crucial for evaluating the trajectory towards fault tolerance and
utility scale. The primary aim of this paper is therefore to present and
analyze key performance metrics derived from experimental realizations
of both physical qubit and QEC implementations. To provide the necessary
context for interpreting these metrics, we first review the essential
concepts at the physical qubit level for various platforms and their
associated metrics in Section~\ref{sec-phys} followed by an overview of
QEC principles and experimental code implementations at the logical
level in Section~\ref{sec-log} analyzing trends in complexity, and the
platforms used.

Given the rapid pace of experimental advancement in the field, tracking
progress requires dynamic tools. To facilitate ongoing tracking and
analysis of this rapid progress, we have developed an open-source
repository containing the databases of experimental results underpinning
this review. This resource, available at
\url{https://github.com/francois-marie/awesome-quantum-computing-experiments},
is designed to be easily updated by the community with new publications
\cite{leregentAwesomeQuantumComputing2025}. Adding new experimental
realizations allows for the automatic regeneration of the figures
presented herein, providing a continuously evolving snapshot of the
state-of-the-art in physical qubit performance and QEC implementation
across different platforms and codes.

Therefore, this work provides the following three key advances: (1) A
unified benchmarking framework aggregating physical and logical qubit
metrics across platforms, (2) Characterization of platform-specific
improvement trends through analysis of 25-year experimental data, and
(3) Open-source repository enabling tracking of documented quantum
experiments through community contributions.

\section{Physical Qubit Benchmarks}\label{sec-phys}

This section delves into the fundamental building blocks of quantum
processors: the physical qubits. As outlined by Nielsen and Chuang
\cite{nielsenQuantumComputationQuantum2000}, the physical realization of
quantum computation necessitates fulfilling four fundamental
requirements: robustly representing quantum information, performing a
universal family of unitary transformations, state preparation, and
measurement. Within this context, critical metrics for assessing
performance include the robustness of quantum states against
environmental noise, often quantified by \(T_1\) and \(T_2\) coherence
times, and the precision in implementing quantum operations,
particularly concerning the controllability required to generate and
manipulate entangled states accurately.

We will thus examine key benchmarks used to assess the quality and
progress of qubits across various leading experimental platforms
\cite{kjaergaardSuperconductingQubitsCurrent2020,
bruzewiczTrappedIonQuantumComputing2019,
winterspergerNeutralAtomQuantum2023,
lauchtRoadmapQuantumNanotechnologies2021,
pezzagnaQuantumComputerBased2021}. We begin by outlining the platforms
considered and the basic concept of how quantum information is encoded
within these physical systems. We will then analyze the physical-level
metrics that quantify performance, including qubit stability via \(T_1\)
and \(T_2\) times, the fidelity of multi-qubit operations through
entanglement error, and the progress towards large-scale systems
indicated by physical qubit counts.

\subsection{Physical Platforms and Qubit Encoding}\label{sec-platform}

To provide essential context for the analysis of performance metrics in
this section, we briefly outline the physical encoding of quantum
information within the hardware platforms considered. Our focus targets
platforms demonstrating significant experimental progress towards QEC
implementation. These include trapped ion systems
\cite{bruzewiczTrappedIonQuantumComputing2019, brownMaterialsChallengesTrappedIon2020},
various superconducting circuits
\cite{kjaergaardSuperconductingQubitsCurrent2020, 
joshiQuantumInformationProcessing2021}, neutral atoms held in dipole
traps \cite{winterspergerNeutralAtomQuantum2023}, Nitrogen vacancy (NV)
centers \cite{pezzagnaQuantumComputerBased2021} and semiconductors
\cite{lauchtRoadmapQuantumNanotechnologies2021}.

Quantum information is fundamentally encoded in the state of two-level
quantum systems, known as qubits. The choice of which two levels within
a larger physical system's Hilbert space represent the computational
basis states \(\ket{0}\) and \(\ket{1}\) varies significantly across
different hardware platforms, each choice presenting distinct advantages
and challenges concerning qubit stability, control fidelity, readout
efficiency, and scalability potential.

For instance, both trapped ions
\cite{bruzewiczTrappedIonQuantumComputing2019, brownMaterialsChallengesTrappedIon2020}
and neutral atoms held in dipole traps
\cite{winterspergerNeutralAtomQuantum2023} commonly encode qubits in
specific internal electronic energy levels. Often, two stable hyperfine
ground states are selected, leveraging their long intrinsic lifetimes.
Manipulation in trapped ions, which uses specific ionic species,
typically involves precisely tuned microwave fields or lasers (depending
on whether hyperfine or optical transitions are used). Neutral atoms,
trapped using laser light (e.g., optical tweezers), also use laser or
microwave fields for control, frequently employing highly excited
Rydberg states to mediate strong interactions required for multi-qubit
gates. While both platforms leverage atomic energy levels, the charged
nature of ions allows for strong confinement via electric fields,
whereas neutral atoms rely on trapping via induced dipole moments which
can be accomplished with optical potentials.

Distinctly, superconducting circuits
\cite{kjaergaardSuperconductingQubitsCurrent2020, joshiQuantumInformationProcessing2021}
realize qubits not within natural atoms, but as the two lowest quantized
energy eigenstates of macroscopic electrical circuits incorporating
non-linear elements like Josephson junctions. These engineered
``artificial atoms'' (e.g., transmons, fluxoniums) have customizable
energy level structures, with \(\ket{0}\) and \(\ket{1}\) corresponding
to distinct energy levels manipulated primarily by microwave pulses.
More complex encodings using higher energy levels or multiple circuit
modes (bosonic codes) offer alternative pathways within this platform.

Another major approach displayed in the metrics of this section relies
on spin degrees of freedom. NV centers
\cite{pezzagnaQuantumComputerBased2021}, which are point defects within
a diamond lattice, typically encode the qubit in the electron spin
ground state, and control is achieved via microwave fields. The inherent
solid-state environment also allows nearby nuclear spins, such as those
of the Nitrogen atom forming the NV center itself (typically \(^{14}\)N
or \(^{15}\)N) or nearby carbon-13 atoms (\(^{13}\)C) naturally present
within the diamond lattice, to serve as qubits or long-lived memories,
often offering longer coherence times than the NV electron spin.
Similarly leveraging spin in a solid-state environment, semiconductor
qubits \cite{lauchtRoadmapQuantumNanotechnologies2021} typically encode
information in the spin state (spin-up \(\ket{\uparrow}\) representing
\(\ket{0}\) and spin-down \(\ket{\downarrow}\) representing \(\ket{1}\),
or vice-versa) of individual charge carriers (electrons or holes)
confined in nanoscale structures like quantum dots, or alternatively,
using the nuclear or electron spin of donor atoms (e.g., phosphorus)
implanted in a host material like silicon. While both NV centers and
semiconductor qubits utilize spin, they differ significantly in their
host material and the specific nature of the spin carrier (defect
electron spin vs. confined electron/hole or donor spin).

As we will see in the next section, each encoding strategy profoundly
influences the qubit's characteristics, including its susceptibility to
different noise sources.

\subsection{Physical Level Metrics}\label{sec-phys-metrics}

Evaluating the hardware platforms requires quantifiable benchmarks. We
now detail several key metrics commonly used to characterize the quality
of individual physical qubits and the scale of experimental systems.

\subsubsection{Bitflip and Coherence times}\label{sec-bit-coh-times}

\begin{figure}[t]

\centering{

\pandocbounded{\includegraphics[keepaspectratio]{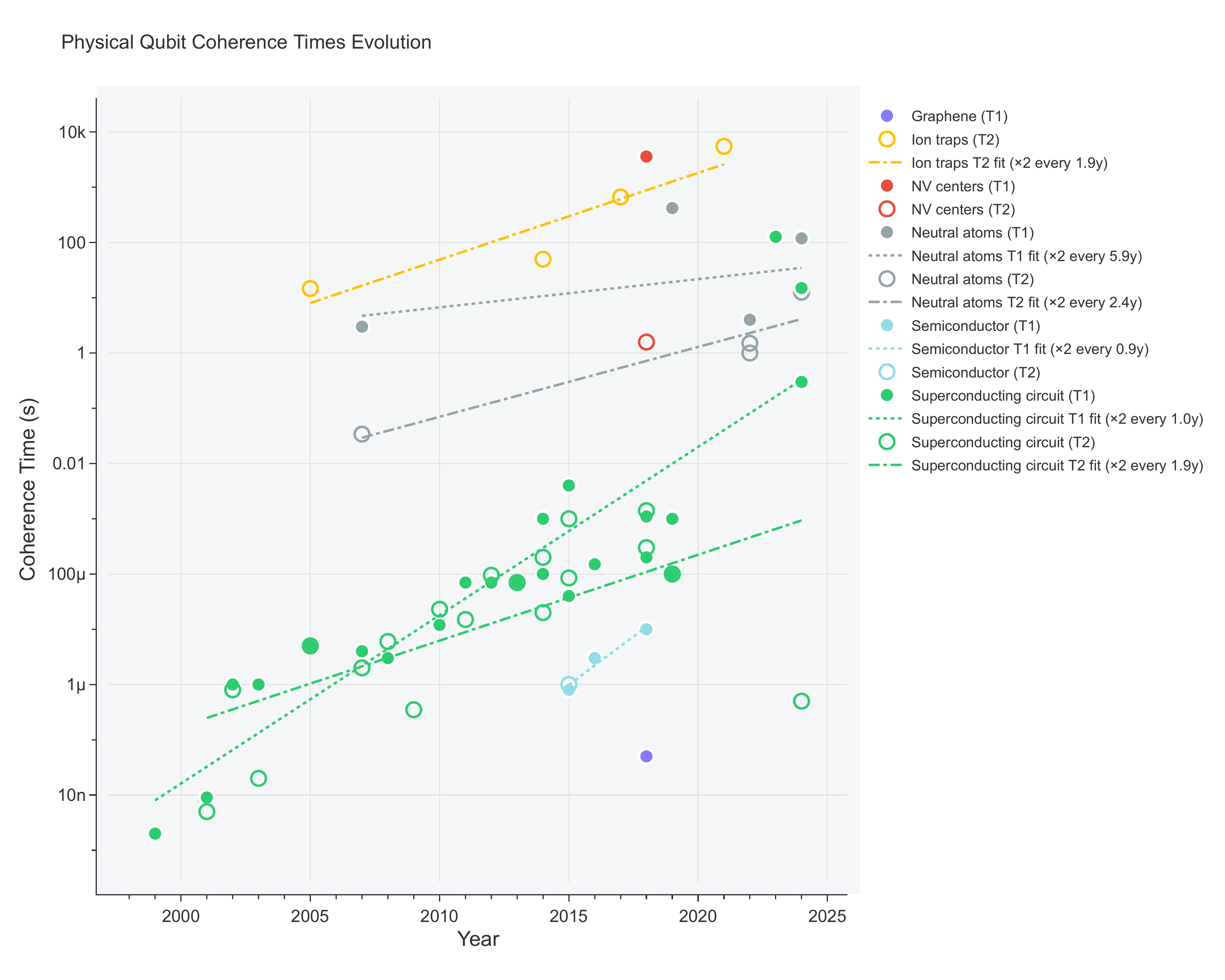}}

}

\caption{\label{fig-cohe}\textbf{Physical qubit bitflip and coherence
times.} Evolution of reported \(T_1\) (solid lines) and \(T_2\) (dashed
lines) times across different platforms. An exponential fit showing the
characteristic doubling time in years is included. The platforms,
ordered by their doubling time (fastest first), are: Semiconductor
\(T_1\) (0.9y but only three data points are displayed), Superconducting
circuit \(T_1\) (1.0y), Ion traps \(T_2\) (1.9y), Superconducting
circuit \(T_2\) (1.9y), Neutral atoms \(T_2\) (2.4y), and Neutral atoms
\(T_1\) (5.9y). See Section~\ref{sec-fit} for additional details on the
fits.}

\end{figure}%

The stability of a physical qubit is fundamentally limited by two
characteristic times: the energy relaxation time (\(T_1\), or bitflip
time) and the dephasing time (\(T_2\), or coherence time). \(T_1\)
quantifies the timescale over which the system incoherently exchanges
energy with its environment. However, the practical factor limiting this
timescale can vary by platform. For example, in neutral atom qubits
where the intrinsic decay between the hyperfine qubit states used is
often negligible, the relevant timescale analogous to \(T_1\) is
typically determined by off-resonant scattering of trapping light rather
than spontaneous emission between \(\ket{1}\) and \(\ket{0}\).
Nevertheless, for the purpose of comparative analysis in this review,
such dominant qubit loss timescales are generally categorized and
plotted under the \(T_1\) metric. \(T_2\) measures the timescale over
which the relative phase between \(\ket{0}\) and \(\ket{1}\) in a
superposition state is lost. Long \(T_1\) and \(T_2\) times are
essential for preserving quantum information.

The
\href{https://github.com/francois-marie/awesome-quantum-computing-experiments/blob/main/data/physical_qubits.csv}{\texttt{physical\_qubits}}
dataset \cite{leregentAwesomeQuantumComputing2025} compiles this key
information on physical qubits developed from 1999 to 2024, documenting
their evolution across various platforms. For each experiment, the
dataset records the code name, the corresponding research article title
and link, the publication year, the platform type, and the coherence
metrics including \(T_1\) and \(T_2\) times. Figure \ref{fig-cohe} plots
this data, illustrating significant improvements over the past decades,
particularly for superconducting circuits
\cite{kjaergaardSuperconductingQubitsCurrent2020}. References are, in
chronological order, for superconducting circuits
\cite{kjaergaardSuperconductingQubitsCurrent2020} based on Josephson
junctions \cite{nakamuraCoherentControlMacroscopic1999a,
nakamuraChargeEchoCooperPair2002,
vionManipulatingQuantumState2002,
chiorescuCoherentQuantumDynamics2003,
bertetDephasingSuperconductingQubit2005,
houckControllingSpontaneousEmission2008,
wangMeasurementDecayFock2008,
manucharyanFluxoniumSingleCooper2009,
bylanderDynamicalDecouplingNoise2011,
paikObservationHighCoherence2011,
rigettiSuperconductingQubitWaveguide2012,
changImprovedSuperconductingQubit2013,
popCoherentSuppressionElectromagnetic2014,
jinThermalResidualExcitedState2015,
ofekExtendingLifetimeQuantum2016,
wangSchrodingerCatLiving2016,
yanFluxQubitRevisited2016,
rosenblumFaulttolerantDetectionQuantum2018,
huDemonstrationQuantumError2019,
nguyenHighcoherenceFluxoniumQubit2019}, and based on bosonic encodings
\cite{wangMeasurementDecayFock2008,
ofekExtendingLifetimeQuantum2016,
wangSchrodingerCatLiving2016,
rosenblumFaulttolerantDetectionQuantum2018,
huDemonstrationQuantumError2019,
lescanneExponentialSuppressionBitflips2020,
berdouOneHundredSecond2023,
marquetAutoparametricResonanceExtending2023,
regladeQuantumControlCatqubit2023}, for semiconductors
\cite{larsenSemiconductorNanowireBasedSuperconducting2015,
casparisGatemonBenchmarkingTwoQubit2016,
luthiEvolutionNanowireTransmons2018}, for trapped ions
\cite{langerLongLivedQubitMemory2005,
hartyHighFidelityPreparationGates2014,
wangSinglequbitQuantumMemory2017,
wangSingleIonQubit2021}, for neutral atoms
\cite{jonesFastQuantumState2007, 
covey2000TimesRepeatedImaging2019,
bluvsteinQuantumProcessorBased2022, 
grahamDemonstrationMultiqubitEntanglement2022, 
manetschTweezerArray61002024} and for graphene
\cite{wangQuantumCoherentControl2019}. The last two decades have seen
substantial improvements in physical qubit stability, with typical
coherence times increasing by one or more orders of magnitude in several
leading experimental platforms. Exponential fits to this data (see
Section~\ref{sec-fit} for goodness-of-fit analysis) reveal
characteristic doubling times ranging from approximately 0.9 years for
semiconductor \(T_1\) to around 5.9 years for neutral atom \(T_1\). It
should be noted that for the superconducting circuit platform in
particular, this overall trend analysis combines data from distinct
qubit encodings, such as transmons and those based on bosonic codes,
which inherently possess different characteristics and may exhibit
varied individual scaling trajectories but they are fitted collectively
here for simplicity.

\subsubsection{Entanglement Error}\label{sec-entangled}

\begin{figure}[ht]

\centering{

\pandocbounded{\includegraphics[keepaspectratio]{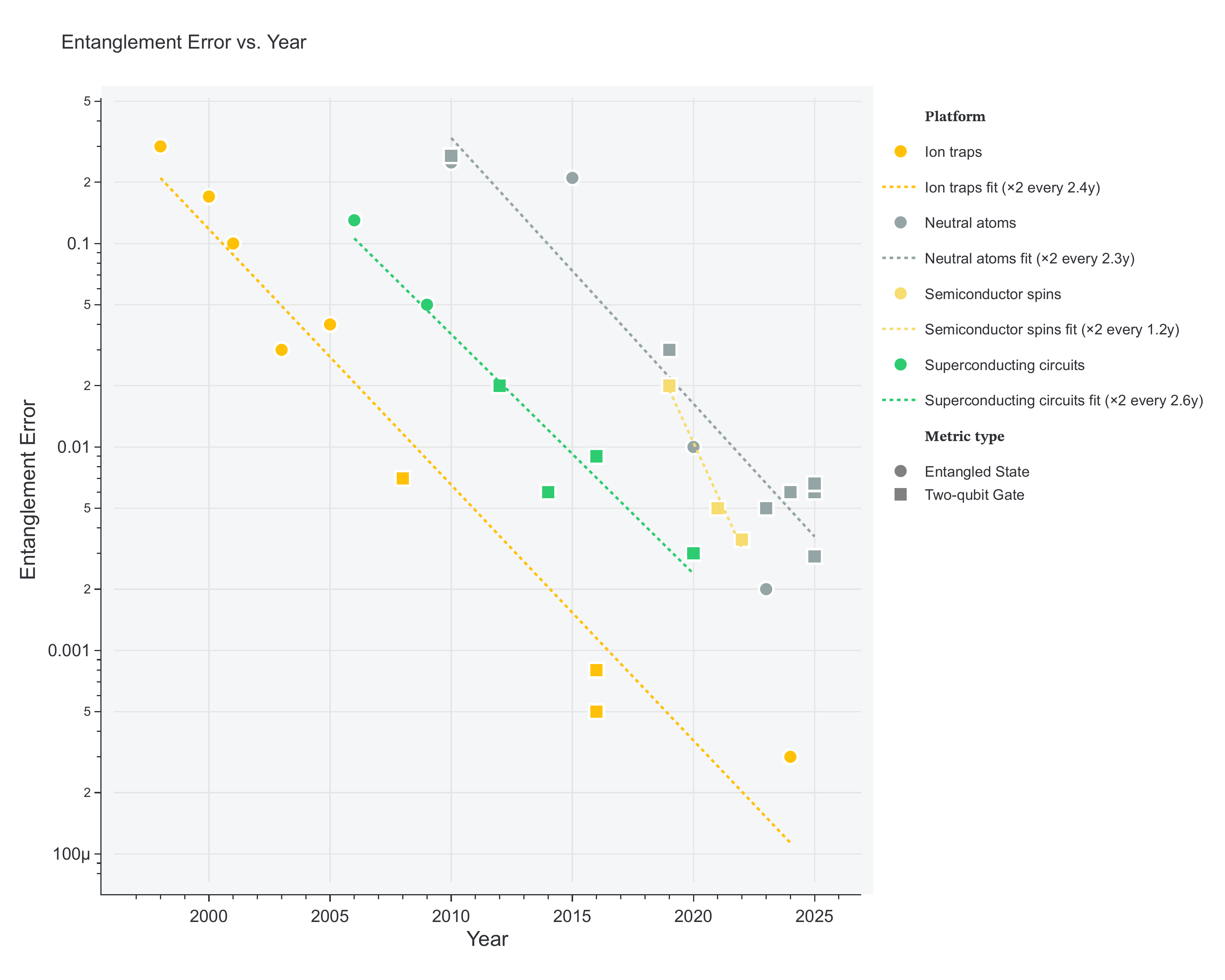}}

}

\caption{\label{fig-enterr}\textbf{Entanglement error.} Data points
represent experimental achievements for ion traps (yellow circles), NV
centers (red circle), neutral atoms (grey circles), semiconductor spins
(light yellow circles), and superconducting circuits (green circles).
Dotted lines show exponential fits to the data for platforms with
sufficient data, indicating the approximate time required to halve the
error rate: Semiconductor spins (1.2y), Neutral atoms (2.3y), Ion traps
(2.4y), and Superconducting circuits (2.6y). See Section~\ref{sec-fit}
for additional details on the fits.}

\end{figure}%

Beyond single-qubit stability, the ability to create high-fidelity
multi-qubit entanglement is paramount for quantum computation and QEC.
For the purpose of this analysis and depending on the available data, we
consider both Bell state preparation error (the error associated with
generating a maximally entangled state) and two-qubit gate error as
equivalent metrics for quantifying entanglement error, as both provide a
measure of the system's ability to create and manipulate entangled
states. The
\href{https://github.com/francois-marie/awesome-quantum-computing-experiments/blob/main/data/entangled_state_error_exp.csv}{\texttt{entangled\_state\_error\_exp}}
dataset tracks the evolution of error rates in entanglement across
multiple quantum computing platforms from 1998 to 2025. For each
experiment, the dataset records the entanglement error rate (entangled
state error or two-qubit gate error), corresponding research article
title, publication year, and implementation platform. Figure
\ref{fig-enterr} plots the reported entanglement error over time for
leading platforms. The references are, in chronological order, for
superconducting circuits \cite{steffenMeasurementEntanglementTwo2006,
dicarloDemonstrationTwoQubitAlgorithms2009,
chowCompleteUniversalQuantum2012,
barendsLogicGatesSurface2014,
sheldonProcedureSystematicallyTuning2016,
kjaergaardProgrammingQuantumComputer2020}, for neutral atoms
\cite{isenhowerDemonstrationNeutralAtom2010,
wilkEntanglementTwoIndividual2010,
mallerRydbergblockadeControllednotGate2015,
levineParallelImplementationHighfidelity2019a,
madjarovHighFidelityEntanglementDetection2020,
schollErasureConversionHighfidelity2023,
everedHighfidelityParallelEntangling2023,
munizHighfidelityUniversalGates2024,
peperSpectroscopyModeling171mathrmYb2025,
tsaiBenchmarkingFidelityResponse2025,
radnaevUniversalNeutralatomQuantum2025}, for trapped ions
\cite{turchetteDeterministicEntanglementTwo1998,
sackettExperimentalEntanglementFour2000,
roweExperimentalViolationBells2001,
leibfriedExperimentalDemonstrationRobust2003a,
haffnerRobustEntanglement2005,
benhelmFaulttolerantQuantumComputing2008,
ballanceHighfidelityQuantumLogic2016,
gaeblerHighFidelityUniversalGate2016,
loschnauerScalableHighfidelityAllelectronic2024}, for semiconductor
spins \cite{huangFidelityBenchmarksTwoqubit2019,
noiriFastUniversalQuantum2022, 
xueQuantumLogicSpin2022} and for NV center
\cite{yamamotoStronglyCoupledDiamond2013}. We observe a consistent trend
of decreasing entanglement error across ion traps, neutral atoms, and
superconducting circuits, all below the sub-percent level. Exponential
fits to the data indicate characteristic error halving times ranging
from approximately 1.2 years for semiconductor spins to 2.6 years for
superconducting circuits, reflecting a consistent trend of rapid
improvement over the past years (see Section~\ref{sec-fit} for more
details). However, recent experimental results are above their
respective fits which indicates that progress seems to be recently
slowing down for ion traps, superconducting circuits and neutral atoms.

\subsubsection{Qubit Count for Scale up}\label{sec-qubit-count}

Achieving fault tolerance requires a large number of physical qubits to
encode logical information and perform QEC
\cite{gidneyHowFactor20482021a, omanakuttanThresholdFaulttolerantQuantum2025}.
Scaling the number of individually controllable, high-quality qubits is
therefore a major engineering challenge. The qubit count metric serves
thus as a critical link between physical qubit progress and the
resources available for implementing QEC codes. It is important to note
that for neutral atom platforms, while trapping large numbers of atoms
in optical lattices is relatively straightforward, the reported qubit
counts specifically refer to systems demonstrating essential quantum
computing capabilities such as site-resolved readout and internal-state
manipulation, rather than just the raw number of trapped atoms. The
\href{https://github.com/francois-marie/awesome-quantum-computing-experiments/blob/main/data/qubit_count.csv}{\texttt{qubit\_count}}
dataset displays experiments from 1998 to 2024, tracking the increasing
number of qubits in experimental quantum systems across various
platforms. For each realization, the dataset records the qubit count,
corresponding research article title, publication year, and
implementation platform.

\begin{figure}[ht]

\centering{

\pandocbounded{\includegraphics[keepaspectratio]{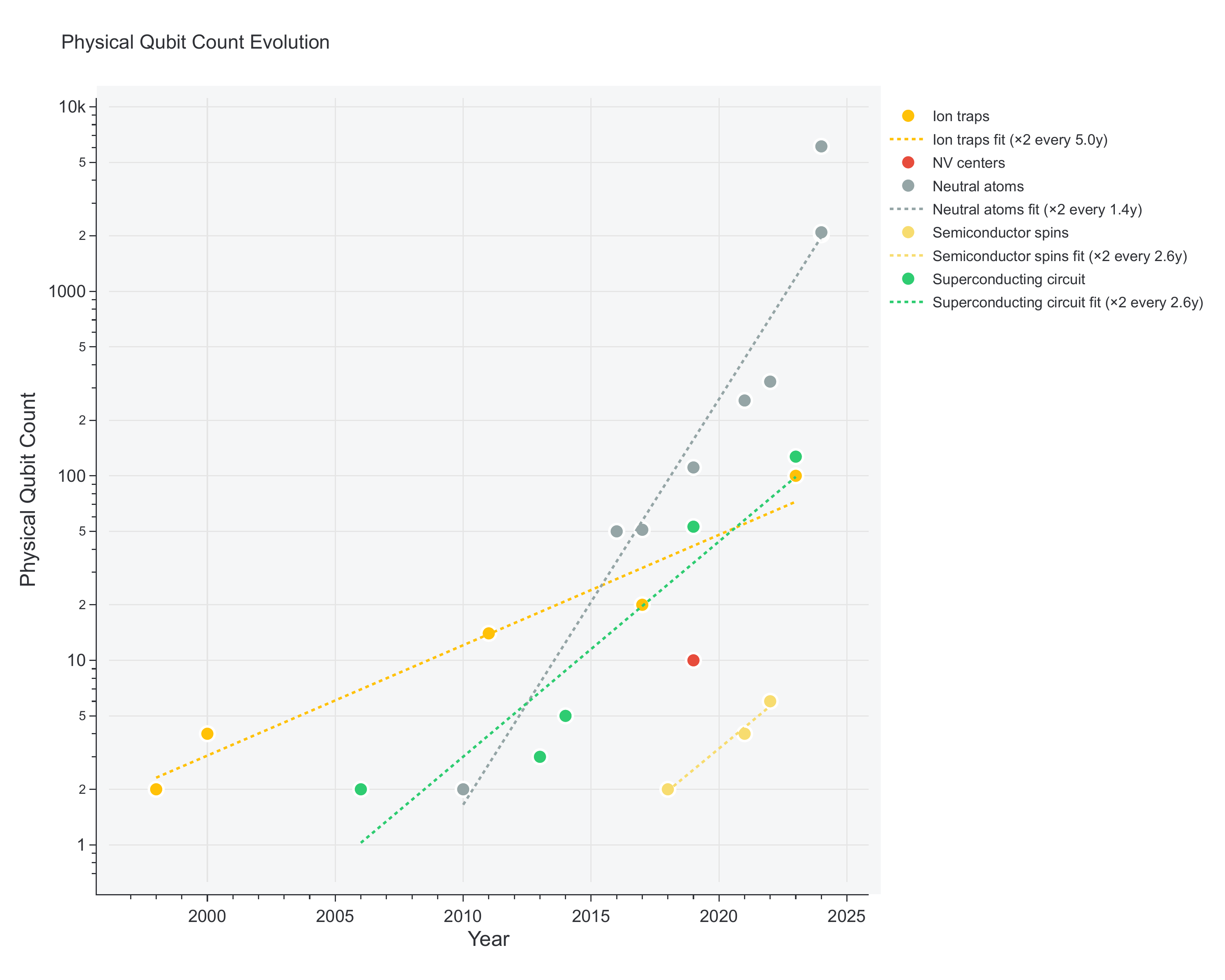}}

}

\caption{\label{fig-qcount}\textbf{Physical qubit count}. Maximum number
of physical qubits reported in experiments over time for selected
platforms. The color code is the same as in Figure~\ref{fig-enterr}.
Dotted lines indicate exponential fits, showing the approximate doubling
time for qubit count: Neutral atoms (1.4y), Superconducting circuits
(2.6y), Semiconductor spins (2.6y), and Ion traps (5.0y). See
Section~\ref{sec-fit} for additional details on the fits.}

\end{figure}%

Figure \ref{fig-qcount} illustrates the growth in the maximum number of
physical qubits reported. The references are, in chronological order,
for ion traps \cite{turchetteDeterministicEntanglementTwo1998,
sackettExperimentalEntanglementFour2000,
monz14qubitEntanglementCreation2011,
friisObservationEntangledStates2018,
kiesenhoferControllingTwodimensionalCoulomb2023}, neutral atoms
\cite{wilkEntanglementTwoIndividual2010,
barredoAtombyatomAssemblerDefectfree2016,
bernienProbingManybodyDynamics2017a,
melloDefectfreeAssembly2D2019,
ebadiQuantumPhasesMatter2021,
schymikInsituEqualizationSingleatom2022,
manetschTweezerArray61002024,
linAIEnabledRapidAssembly2024,
pichardRearrangementIndividualAtoms2024}, superconducting circuit
\cite{steffenMeasurementEntanglementTwo2006,
chowImplementingStrandScalable2014,
barendsSuperconductingQuantumCircuits2014,
aruteQuantumSupremacyUsing2019,
kimEvidenceUtilityQuantum2023}, semiconductor spins
\cite{watsonProgrammableTwoqubitQuantum2018a,
hendrickxFourqubitGermaniumQuantum2021,
philipsUniversalControlSixqubit2022}, and NV centers
\cite{bradleyTenQubitSolidStateSpin2019}. The data highlights
platform-specific scaling trajectories, with neutral atoms demonstrating
the largest qubit counts (6,100 qubits), followed by superconducting
circuits (127 qubits), and ion traps (100 qubits). Fitting exponential
trends to this data reveals different scaling rates, with neutral atoms
exhibiting the fastest approximate doubling time at 1.4 years, compared
to 2.6 years for superconducting circuits and 5.0 years for ion traps.

Section~\ref{sec-phys} detailed the remarkable advancements in
physical-qubit technologies over the past decades, showcasing
multi-order-of-magnitude improvements in coherence times, entanglement
fidelity, and achievable system sizes across various platforms. However,
as the analysis of entanglement error (Section~\ref{sec-entangled})
illustrates, even state-of-the-art physical qubits possess inherent
error rates and face scaling challenges that currently preclude the
direct execution of large-scale, fault-tolerant quantum algorithms. The
physical error rates, while dramatically reduced over the past decades,
remain significantly higher than what complex computations demand, and
bridging this fidelity gap is the primary motivation for QEC. Therefore,
translating the physical qubit advancements reviewed previously into
reliable computational power depends critically on the effective
implementation and scaling of QEC protocols. Therefore, we now
transition from the physical layer benchmarks to the logical layer, by
first delving into the principles of QEC and then surveying the
experimental landscape of QEC code implementations designed to protect
quantum information and ultimately enable fault-tolerant computation.

\section{QEC Code Implementation}\label{sec-log}

Inspired by Gottesman's prioritization of specific QEC codes based on
their resource requirements for early fault-tolerance demonstrations
\cite{gottesmanQuantumFaultTolerance2016}, numerous experimental efforts
have successfully implemented many of these codes, see
Table~\ref{tbl-early}. These realizations affirm the viability of using
smaller, experimentally accessible codes as the initial steps towards
achieving robust, fault-tolerant quantum computation.

\begin{longtable}[]{@{}
  >{\centering\arraybackslash}p{(\linewidth - 4\tabcolsep) * \real{0.5497}}
  >{\centering\arraybackslash}p{(\linewidth - 4\tabcolsep) * \real{0.2649}}
  >{\centering\arraybackslash}p{(\linewidth - 4\tabcolsep) * \real{0.1854}}@{}}
\caption{\textbf{Minimum Qubits to demonstrate Fault Tolerance on early
implementation} \cite{gottesmanQuantumFaultTolerance2016}. The 7-qubit
code requires 4 ancilla for Error Correction and 1 for testing. We make
separate rows for the standard 9-qubit Bacon-Shor code and the
two-dimensional (2D) nearest-neighbor (NN)
implementation.}\label{tbl-early}\tabularnewline
\toprule\noalign{}
\begin{minipage}[b]{\linewidth}\centering
Quantum Code Name
\end{minipage} & \begin{minipage}[b]{\linewidth}\centering
Minimum Qubits for Fault Tolerance
\end{minipage} & \begin{minipage}[b]{\linewidth}\centering
Number of Ancilla Qubits
\end{minipage} \\
\midrule\noalign{}
\endfirsthead
\toprule\noalign{}
\begin{minipage}[b]{\linewidth}\centering
Quantum Code Name
\end{minipage} & \begin{minipage}[b]{\linewidth}\centering
Minimum Qubits for Fault Tolerance
\end{minipage} & \begin{minipage}[b]{\linewidth}\centering
Number of Ancilla Qubits
\end{minipage} \\
\midrule\noalign{}
\endhead
\bottomrule\noalign{}
\endlastfoot
\([[4, 2, 2]]\) code & 5 & 1 \\
\([[6, 4, 2]]\) code & 7 & 1 \\
\([[8, 6, 2]]\) code & 9 & 1 \\
9-qubit Bacon-Shor code & 10 & 1 \\
5-qubit code
\cite{laflammePerfectQuantumError1996, bennettMixedStateEntanglement1996}
& 11 & 6 \\
7-qubit code \cite{calderbankGoodQuantumErrorCorrecting1996} & 12 & 5 \\
Surface code & \(\geq\) 13 & - \\
2D NN 9-qubit Bacon-Shor code & 13 & 1 \\
\end{longtable}

Building upon these early demonstrations, this section provides a
detailed examination of the broader landscape of experimental QEC code
implementations. We will explore the underlying concepts, survey the
range of codes and platforms utilized, and analyze key metrics tracking
progress towards fault tolerance.

\subsection{Concept of QEC}\label{sec-qec-concept}

QEC provides a strategy to protect fragile quantum information from
noise inherent in physical systems and gate operations. The core
principle involves encoding the state of one or more logical qubits into
a carefully chosen subspace within a larger Hilbert space spanned by
many physical qubits. QEC codes are specifically designed such that
common types of physical errors (e.g., single-qubit bit flips or phase
flips) map the encoded logical state onto distinguishable, mutually
orthogonal error subspaces. This structure enables to extract
information about the error that occurred, called error syndrome,
through measurements, often involving ancillary qubits. This process
reveals the error type and location without directly measuring, and thus
collapsing, the encoded logical information itself, allowing for
subsequent correction.

The fundamental properties and capabilities of a QEC code are concisely
described by the parameters \([[n, k, d]]\). Here, \(n\) represents the
total number of physical qubits utilized in the code block, \(k\)
denotes the number of logical qubits robustly encoded, and \(d\) is the
code distance. The distance \(d\) is a crucial metric quantifying the
code's error detection and correction power. It corresponds to the
minimum number of arbitrary single-qubit errors that could potentially
transform one encoded logical state into another, or cause an
undetectable error. A code possessing distance \(d\) is guaranteed to
detect any combination of up to \(d-1\) physical errors and correct any
combination of up to \(\lfloor (d-1)/2 \rfloor\) physical errors
occurring within a code block. Consequently, achieving higher code
distances offers enhanced protection against noise, but this typically
comes at the cost of requiring a significantly larger number of physical
qubits \(n\).

With the fundamental concepts of encoding, error detection, and code
parameters \([[n,k,d]]\) established, we now survey the key experimental
efforts realizing these principles. This review concentrates on the
platforms and specific QEC codes where significant progress has been
documented in the scientific literature, including both peer-reviewed
publications and widely accessible preprints (e.g., via arXiv),
providing the basis for the analysis presented in the following
sections.

\subsection{Experiments Considered}\label{sec-qec-codes}

Our analysis focuses on the platforms that have demonstrated the most
significant experimental progress in multi-qubit operations and the
implementation of QEC. These include established approaches like trapped
ions
\cite{bruzewiczTrappedIonQuantumComputing2019, brownMaterialsChallengesTrappedIon2020}
and various superconducting circuits
\cite{kjaergaardSuperconductingQubitsCurrent2020, joshiQuantumInformationProcessing2021},
alongside rapidly advancing systems such as arrays of neutral atoms held
in optical traps \cite{winterspergerNeutralAtomQuantum2023}. We also
consider photonic systems \cite{knillSchemeEfficientQuantum2001}, which
often leverage different computational models based on linear optics or
integrated circuits, and solid-state spin systems like NV centers in
diamond \cite{pezzagnaQuantumComputerBased2021}. For historical context
and early demonstrations, results from Nuclear Magnetic Resonance (NMR)
are also included.

The experimental exploration of QEC spans a wide spectrum of codes,
reflecting different strategies and complexities. Earliest work often
implemented simple Repetition Codes
\cite{albertQuantumRepetitionCode2022} (distance \(d\), realizing
\([[d, 1, d]]\) parameters for either bit-flip or phase-flip protection)
as building blocks or initial demonstrations
\cite{coryExperimentalQuantumError1998,
chiaveriniRealizationQuantumError2004,
moussaDemonstrationSufficientControl2011,
reedRealizationThreeQubitQuantum2011,
schindlerExperimentalRepetitiveQuantum2011,
zhangExperimentalQuantumError2011,
waldherrQuantumErrorCorrection2014,
kellyStatePreservationRepetitive2015,
risteDetectingBitflipErrors2015,
cramerRepeatedQuantumError2016,
woottonRepetitionCode152018,
woottonBenchmarkingNeartermDevices2020,
chenExponentialSuppressionBit2021,
acharyaSuppressingQuantumErrors2022,
acharyaQuantumErrorCorrection2024,
puttermanHardwareefficientQuantumError2025}. Significant effort has also
focused on early stabilizer codes like the \([[5, 1, 3]]\) Perfect Code
\cite{knillImplementationFiveQubit2001,
zhangExperimentalImplementationEncoded2012,
gongExperimentalExplorationFivequbit2022}, as well as various small
error detecting codes like the \([[4, 1, 2]]\) or \([[4, 2, 2]]\)
variants \cite{linkeFaulttolerantQuantumError2017,
takitaExperimentalDemonstrationFaultTolerant2017,
roffeProtectingQuantumMemories2018,
willschTestingQuantumFault2018,
vuillotErrorDetectionHelpful2018,
harperFaultTolerantLogicalGates2019a,
koleResourceOptimalRealization2020,
sunOpticalDemonstrationQuantum2020,
urbanekErrorDetectionQuantum2020,
caneExperimentalCharacterizationFaultTolerant2021,
chenExponentialSuppressionBit2021,
zhangComparativeAnalysisError2022,
bedalovFaultTolerantOperationMaterials2024,
damEndtoEndQuantumSimulation2024,
guptaEncodingMagicState2024,
reichardtLogicalComputationDemonstrated2024}. More recent research
increasingly targets codes with topological properties or higher
potential for fault tolerance, such as Surface Codes
\cite{albertRotatedSurfaceCode2024} implemented with various distances
and configurations, often encoding \(k=1\)
\cite{bellExperimentalDemonstrationGraph2014,
andersenRepeatedQuantumError2020,
acharyaSuppressingQuantumErrors2022,
bluvsteinQuantumProcessorBased2022,
krinnerRealizingRepeatedQuantum2022,
acharyaQuantumErrorCorrection2024,
bluvsteinLogicalQuantumProcessor2024,
eickbuschDemonstratingDynamicSurface2024}, or \(k=2\)
\cite{bluvsteinQuantumProcessorBased2022} logical qubits, and Color
Codes (e.g., \([[7, 1, 3]]\), \([[19, 1, 5]]\), potentially encoding
\(k=1, 2,\) or even \(3\) logical qubits)
\cite{niggExperimentalQuantumComputations2014,
bluvsteinQuantumProcessorBased2022,
bluvsteinLogicalQuantumProcessor2024,
lacroixScalingLogicColor2024,
rodriguezExperimentalDemonstrationLogical2024}. Subsystem codes, notably
the Bacon-Shor code
\cite{shorSchemeReducingDecoherence1995, baconDecoherenceControlSymmetry2003, baconOperatorQuantumError2006}
(e.g., \([[9, 1, 3]]\)) \cite{eganFaultTolerantOperationQuantum2021,
luoQuantumTeleportationPhysical2021,
reichardtLogicalComputationDemonstrated2024}, offer alternative
structures for fault tolerance. Additionally, numerous demonstrations
use simpler entangled states like Bell states
\cite{corcolesDemonstrationQuantumError2015a,
andersenEntanglementStabilizationUsing2019,
bultinkProtectingQuantumEntanglement2020}, primarily for error detection
protocols rather than full correction, or employ Cluster States
\cite{bluvsteinQuantumProcessorBased2022},~ as resources for
measurement-based quantum computation which also possess inherent error
detection properties. The specific implementations of these codes across
the aforementioned platforms form the core dataset analyzed in this
section.

Beyond identifying the specific platforms and codes explored,
quantifying the progress in QEC experiments requires analyzing dedicated
metrics related to these implementations.

\subsection{Logical Level Metrics}\label{sec-qec-metrics}

Evaluating the progress in experimental QEC necessitates to track
specific metrics that capture the type of code implemented, the hardware
platform, the \([[n, k, d]]\) code parameters, and the achieved
performance, often assessed through logical error rates and proximity to
fault-tolerance accuracy thresholds. The analysis presented in this
section is based on the
\href{https://github.com/francois-marie/awesome-quantum-computing-experiments/blob/main/data/qec_exp.csv}{\texttt{qec\_exp}}
dataset, which compiles information on experimental QEC implementations
spanning from the earliest demonstrations in 1998
\cite{coryExperimentalQuantumError1998} up to recent results in 2024.
This dataset covers a diverse range of platforms mentioned above,
including superconducting circuits, trapped ions, neutral atoms, NMR
systems, and photonic devices.

For each documented implementation, the dataset records key details: the
specific QEC code family (e.g., Repetition Code, Surface Code, Color
Code), the corresponding publication reference, year of publication, the
code's \([[n, k, d]]\) parameters, and the hardware platform employed.
Examining the data chronologically reveals a clear trajectory of
progress: beginning with three-qubit repetition codes primarily in NMR
systems, the field has advanced significantly to implementing
sophisticated codes like surface codes that leverage up to 49 physical
qubits on neutral atom arrays and superconducting circuits. Notably,
recent milestones captured in the dataset include demonstrations of
error correction operating below the fault-tolerance threshold using
surface codes \cite{acharyaQuantumErrorCorrection2024}.

\subsubsection{Cumulative Growth and Platform
Distribution}\label{sec-qec-platform}

\begin{figure}[ht]

\centering{

\pandocbounded{\includegraphics[keepaspectratio]{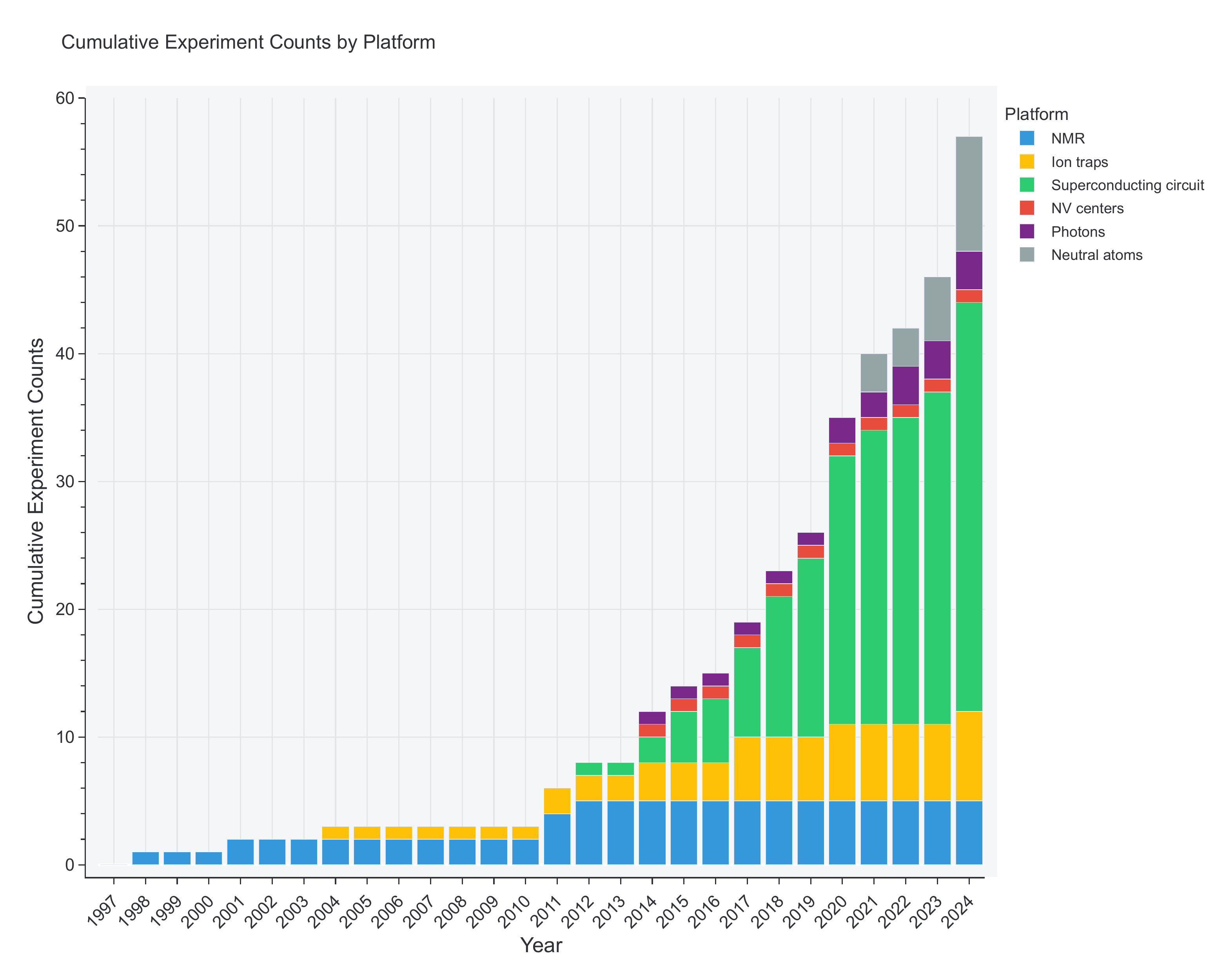}}

}

\caption{\label{fig-exp-cumul}\textbf{Cumulative experiment counts by
platform.} Total number of published QEC-related experiments up to a
given year, categorized by platform. See Section~\ref{sec-yearly} for a
yearly (non cumulative) representation}

\end{figure}%

The field has witnessed substantial growth in experimental QEC activity
over the past quarter-century. Figure \ref{fig-exp-cumul} plots the
cumulative count of documented QEC experiments since 1998, broken down
by the physical platform employed. The overall trajectory demonstrates
significant and accelerating growth in research activity. While NMR
platforms hosted the earliest demonstrations, trapped ion systems began
contributing consistently around 2004 and show steady growth.
Superconducting circuits emerged shortly thereafter and have since
become the dominant platform in terms of cumulative experimental output,
exhibiting particularly rapid growth in the last decade. Neutral atom
platforms represent a more recent but fast-growing contributor to the
field, alongside consistent contributions from photonic systems and
nascent efforts in NV centers. This cumulative view highlights the
increasing investment and progress across multiple hardware modalities
in tackling the challenges of QEC. A non-cumulative, year-by-year
breakdown is provided in Figure \ref{fig-exp-yearly} in the Appendices.

\subsubsection{\texorpdfstring{\([[n, k, d]]\)
Representation}{{[}{[}n, k, d{]}{]} Representation}}\label{sec-qec-nkd}

\begin{figure}[ht]

\centering{

\pandocbounded{\includegraphics[keepaspectratio]{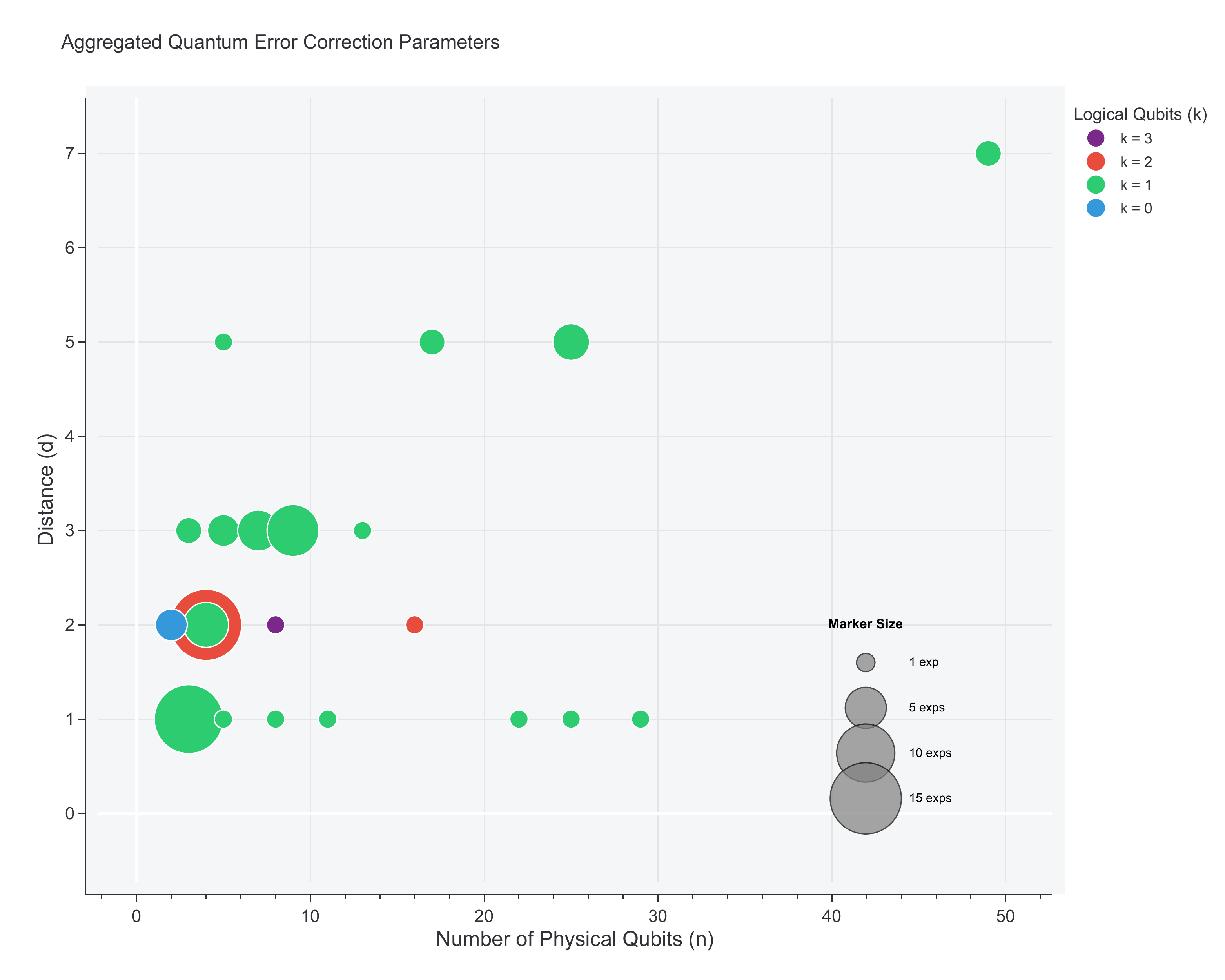}}

}

\caption{\label{fig-nkd}\textbf{QEC Code Parameters \([[n, k, d]]\).}
Scatter plot showing the number of physical qubits (\(n\)) versus the
code distance (\(d\)) for experimentally implemented QEC codes. Points
are labeled by the code type and logical qubit count (\(k\)). We see the
horizontal line at \(d=1\) corresponding to the classical repetition
code and the square root profile of the surface code where
\(d \propto \sqrt{n}\). Realized codes focus on encoding a single
logical qubit (\(k=1\)), even though implementations with more logical
qubits do exist.}

\end{figure}%

As mentioned above, the \([[n, k, d]]\) parameters provide a concise
summary of a QEC code's resource requirements (\(n\)), encoded
information capacity (\(k\)), and error correction strength (\(d\)).
Figure \ref{fig-nkd} visualizes the landscape of experimentally realized
QEC codes by plotting the number of physical qubits (\(n\)) against the
achieved code distance (\(d\)). Each point represents a documented
experiment, labeled by the code type and the number of logical qubits
encoded (\(k\)).

The plot reveals that a significant portion of experimental work has
focused on codes with relatively low distance (\(d \le 3\)) and using
fewer than 50 physical qubits. These implementations often serve as
crucial demonstrations of QEC principles, stabilizer measurements, or
specific code components. Notably, simple repetition codes (\(k=1\),
various \(d\)) and early stabilizer codes like the \([[5,1,3]]\) code
are represented in this region. However, the figure also clearly
illustrates the ongoing effort to push towards higher code distances,
which offer protection against a greater number of physical errors,
which is a prerequisite for fault tolerance. Milestones include
implementations of surface codes and color codes reaching \(d=3\),
\(d=5\) and even \(d=7\), requiring substantially more physical qubits
(approaching \(n=50\) for \(d=7\)).

\subsubsection{QEC Performance
Benchmarks}\label{qec-performance-benchmarks}

While Section~\ref{sec-qec-metrics} introduces logical level metrics
such as the cumulative count of experiments and the \([[n, k, d]]\)
parameters realized, a dedicated analysis of the performance of
implemented QEC codes is essential for benchmarking progress towards
fault tolerance. Key performance indicators include the suppression of
logical errors relative to physical errors, often quantified by the
suppression factor \(\Lambda\), demonstrations of logical qubit lifetime
extension beyond the physical qubit limit (break-even), and operation
below theoretical fault-tolerance thresholds. Recent experiments have
made significant strides in these areas. For instance, logical error
suppression when increasing code distance has been quantified, scaling
from distance-3 to distance-5, suppression factors of approximately
\(2.15(2)\), \(1.69(6)\), and \(1.56(2)\) were measured for different
dynamic surface code implementations
\cite{eickbuschDemonstratingDynamicSurface2024}, and a factor of
\(\Lambda = 1.56(4)\) was observed for the color code, which simulations
indicated was below threshold \cite{lacroixScalingLogicColor2024}. While
an earlier surface code experiment showed only modest improvement
between distance-3 and distance-5 logical qubits (logical error per
cycle of \(2.914\%\) vs
\(3.028\%\))\cite{acharyaSuppressingQuantumErrors2022}, the following
demonstration achieved significant suppression
(\(\Lambda = 2.14 \pm 0.02\)) when scaling from distance-5 to distance-7
\cite{acharyaQuantumErrorCorrection2024}. This latter work also reported
operating both distance-5 and distance-7 surface codes below the
fault-tolerance threshold, achieving a logical error rate per cycle of
\(0.143\% \pm 0.003\%\) for the 101-qubit distance-7 code
\cite{acharyaQuantumErrorCorrection2024}. Furthermore, this distance-7
logical memory surpassed the break-even point, exhibiting a lifetime
\(2.4 \pm 0.3\) times longer than its best constituent physical qubit
\cite{acharyaQuantumErrorCorrection2024}. Although these recent
achievements mark critical milestones, the overall number of such
demonstrations remains insufficient to establish statistically
significant trends in logical error suppression or performance scaling
across different codes and platforms at this time.

\section{Conclusion}\label{sec-ccl}

In this work, we have introduced an online and open-source tool
\cite{leregentAwesomeQuantumComputing2025} to track the progress of
several chosen metrics in the quantum computing stack, from low-level
characteristics like relaxation and coherence times to higher-level ones
like QEC code parameters.

Our analysis reveals remarkable advances: physical-qubit coherence and gate
fidelity have improved by orders of magnitude, while processor scale has
grown exponentially, enabling access to hundreds and even thousands of
qubits
\cite{manetschTweezerArray61002024, kimEvidenceUtilityQuantum2023}.
Concurrently, experimental QEC has progressed from foundational
demonstrations
\cite{coryExperimentalQuantumError1998, chiaveriniRealizationQuantumError2004}
to implementing sophisticated codes like surface and color codes with
increasing scale \(n\) and distance \(d\) across multiple leading
platforms
\cite{acharyaQuantumErrorCorrection2024, krinnerRealizingRepeatedQuantum2022, bluvsteinLogicalQuantumProcessor2024}.

Despite the impressive progress, achieving robust fault-tolerant quantum
computation requires overcoming persistent critical challenges. Three
key obstacles include: (1) The resource overhead problem: QEC requires
encoding information across multiple physical qubits, dramatically
increasing hardware complexity. For instance, factorizing a 2048-bit RSA
integer using Shor's algorithm would require approximately one million
superconducting qubits despite only needing 1,537 algorithmic qubits,
with each algorithmic qubit requiring at most 1,352 physical qubits for
noise suppression (\(d=25\)) \cite{gidneyHowFactor20482025}, (2) The
processing of encoded information: Logical gates must operate directly
on encoded qubits without decoding, requiring complex physical
operations where a single logical two-qubit gate can involve thousands
of state preparations and physical gate operations. Achieving universal
gate sets adds further overhead through magic state distillation \cite{bravyiUniversalQuantumComputation2005} or
cultivation \cite{gidneyMagicStateCultivation2024}, which can consume a significant fraction of logical qubits
\cite{guptaEncodingMagicState2024, rodriguezExperimentalDemonstrationLogical2024, lacroixScalingLogicColor2024},
and (3) The decoding problem: Decoders must process error syndrome data
at quantum computer clock speeds, potentially handling
gigabit-per-second data rates per logical qubit, requiring scalable
decoding algorithms that balance accuracy, latency, and scalability for
these thousands of logical qubits involved.

While long \(T_1\) and \(T_2\) times are essential for preserving
quantum information, the comparison with gate and QEC cycle times
ultimately determines how many operations can be performed before
quantum information is lost. Future work will thus focus on
incorporating finer-grained metrics essential for fault-tolerance such
as the QEC cycle time but also logical qubit performance under repeated
correction cycles, logical error suppression factors \(\Lambda\),
post-selection rates, detailed ancilla qubit counts, logical qubit
half-lives, alongside benchmarks for key subroutines including magic
state distillation
\cite{guptaEncodingMagicState2024, rodriguezExperimentalDemonstrationLogical2024, lacroixScalingLogicColor2024}
and GHZ state preparation
\cite{mosesRaceTrackTrappedIon2023, baoCreatingControllingGlobal2024, reichardtDemonstrationQuantumComputation2024, bluvsteinLogicalQuantumProcessor2024, hongEntanglingFourLogical2024, reichardtLogicalComputationDemonstrated2024}.
Synthesizing multi-layered progress, facilitated by the presented
resource, is vital for navigating this rapidly evolving field towards
the ultimate goal of impactful fault-tolerant quantum computation.

Designed for flexibility and community contribution, our tool provides a
unified resource that will be developed as the field continues to grow.
We hope it proves valuable for physicists and QEC researchers by having
all the resources in one place to make fair comparisons between
technologies.

\section{Acknowledgements}\label{acknowledgements}

We thank Pascal Scholl and Antoine Browaeys for insightful discussions
during the preparation of this manuscript. We thank Julius de Hond and
Louis Vignoli for reading a draft of this paper and providing useful
feedback. We thank the Pasqal team, and in particular Lucas Lassablière
and Adrien Signoles, for cultivating an environment that enabled this
work.

\section{Appendices}\label{appendices}

\subsection{Reaching the Utility Scale}\label{sec-us}

To contextualize the physical qubit benchmarks within the broader goal
of useful quantum computation \cite{preskillNISQMegaquopMachine2025}, we
project potential timelines towards a `utility scale' regime, employing
a working definition of achieving both a Physical Qubit Count (PQC) of
10,000 and an Entanglement Error (EE) below 0.1\% to be a factor 10
below the threshold of the surface code. These metrics are chosen
somehow arbitrarily as being hard to reach but not impossible in near
future and should be thought of as one possible scenario rather than the
definition of utility. While individual gate errors (such as the
two-qubit gate error) are commonly used to evaluate the distance to the
accuracy threshold in QEC simulations, we use the entangled state error
when it is the only available metric as it benchmarks the system's
capability to produce the high-fidelity entangled states essential for
QEC, providing a functionally relevant and cross-platform comparable
proxy for approaching fault-tolerance requirements. Besides, while one
might specifically track the number of physical qubits directly employed
within QEC code experiments (the focus of Section~\ref{sec-log}),
examining the maximum reported physical qubit count remains a vital
benchmark, as it reflects the underlying platform's overall progress in
fabrication, control, and potential scale, providing the resource pool
from which future, larger QEC implementations must draw. By
extrapolating the observed exponential improvement rates (halving times
for EE, doubling times for PQC) summarized in Table \ref{tbl-us}, we can
estimate when each platform might reach these specific thresholds.

\begin{longtable}[]{@{}
  >{\raggedright\arraybackslash}p{(\linewidth - 16\tabcolsep) * \real{0.1786}}
  >{\centering\arraybackslash}p{(\linewidth - 16\tabcolsep) * \real{0.0714}}
  >{\centering\arraybackslash}p{(\linewidth - 16\tabcolsep) * \real{0.1071}}
  >{\centering\arraybackslash}p{(\linewidth - 16\tabcolsep) * \real{0.1286}}
  >{\centering\arraybackslash}p{(\linewidth - 16\tabcolsep) * \real{0.0857}}
  >{\centering\arraybackslash}p{(\linewidth - 16\tabcolsep) * \real{0.0714}}
  >{\centering\arraybackslash}p{(\linewidth - 16\tabcolsep) * \real{0.1071}}
  >{\centering\arraybackslash}p{(\linewidth - 16\tabcolsep) * \real{0.1643}}
  >{\centering\arraybackslash}p{(\linewidth - 16\tabcolsep) * \real{0.0857}}@{}}
\caption{\textbf{Projected Utility-Scale (US) for Entanglement Error
(EE) and Physical Qubit Count (PQC)}. Best achieved (with corresponding
year), doubling factor and estimated year of reaching the utility scale
by platforms.}\label{tbl-us}\tabularnewline
\toprule\noalign{}
\begin{minipage}[b]{\linewidth}\raggedright
\textbf{Platform}
\end{minipage} & \begin{minipage}[b]{\linewidth}\centering
Best EE
\end{minipage} & \begin{minipage}[b]{\linewidth}\centering
Best EE Year
\end{minipage} & \begin{minipage}[b]{\linewidth}\centering
EE ÷2 factor
\end{minipage} & \begin{minipage}[b]{\linewidth}\centering
\textbf{US EE}
\end{minipage} & \begin{minipage}[b]{\linewidth}\centering
Best PQC
\end{minipage} & \begin{minipage}[b]{\linewidth}\centering
Best PQC Year
\end{minipage} & \begin{minipage}[b]{\linewidth}\centering
PQC ×2 factor
\end{minipage} & \begin{minipage}[b]{\linewidth}\centering
\textbf{US PQC}
\end{minipage} \\
\midrule\noalign{}
\endfirsthead
\toprule\noalign{}
\begin{minipage}[b]{\linewidth}\raggedright
\textbf{Platform}
\end{minipage} & \begin{minipage}[b]{\linewidth}\centering
Best EE
\end{minipage} & \begin{minipage}[b]{\linewidth}\centering
Best EE Year
\end{minipage} & \begin{minipage}[b]{\linewidth}\centering
EE ÷2 factor
\end{minipage} & \begin{minipage}[b]{\linewidth}\centering
\textbf{US EE}
\end{minipage} & \begin{minipage}[b]{\linewidth}\centering
Best PQC
\end{minipage} & \begin{minipage}[b]{\linewidth}\centering
Best PQC Year
\end{minipage} & \begin{minipage}[b]{\linewidth}\centering
PQC ×2 factor
\end{minipage} & \begin{minipage}[b]{\linewidth}\centering
\textbf{US PQC}
\end{minipage} \\
\midrule\noalign{}
\endhead
\bottomrule\noalign{}
\endlastfoot
Ion traps & 0.0003 & 2024 & 2.4 & \textbf{2024} & 100 & 2023 & 5.0 &
\textbf{2056} \\
Neutral atoms & 0.0020 & 2023 & 2.3 & \textbf{2025} & 6,100 & 2024 & 1.4
& \textbf{2024} \\
Superconducting circuits & 0.0030 & 2020 & 2.6 & \textbf{2024} & 127 &
2023 & 2.6 & \textbf{2039} \\
Semiconductor spins & 0.0035 & 2022 & 1.2 & \textbf{2024} & 6 & 2022 &
2.6 & \textbf{2050} \\
\end{longtable}

This analysis projects that neutral atom platforms could satisfy both
criteria as early as 2025, substantially sooner than trapped ions
(projected for 2038) and superconducting circuits (2039). Within the
assumptions of this model, the rapid PQC scaling factor currently
observed for neutral atoms is the primary driver for this accelerated
timeline.

However, several critical caveats apply to these projections. Firstly,
let us emphasize again that the definition of `utility scale' based
solely on PQC and EE is a simplification: true utility will depend on a
complex interplay of metrics including logical qubit fidelity,
connectivity, gate speeds, and the ability to execute fault-tolerant
protocols (as discussed in Section~\ref{sec-log}), not just raw physical
qubit counts and error rates. The 10k and 0.1\% target serves as one
specific, potentially arbitrary, milestone. Secondly, the projection
relies heavily on the assumption that current exponential scaling trends
will continue for years or decades. Such extrapolations are inherently
speculative, as technological progress may encounter unforeseen plateaus
due to fundamental physics or engineering hurdles (as it seems to be
already the case for the EE), or conversely, accelerate due to
breakthroughs not captured by historical data. Factors like increasing
crosstalk at scale, limitations in control systems, or shifts in
research focus could significantly alter these trajectories. Therefore,
while illustrative, these projections should be interpreted cautiously
as reflecting potential timelines if current rates of progress persist
and if the chosen PQC/EE targets adequately capture the requirements for
utility.

\subsection{Additional Visualisations}\label{sec-app-viz}

\subsubsection{Yearly Experiment Counts by Platform}\label{sec-yearly}

\begin{figure}[ht]

\centering{

\pandocbounded{\includegraphics[keepaspectratio]{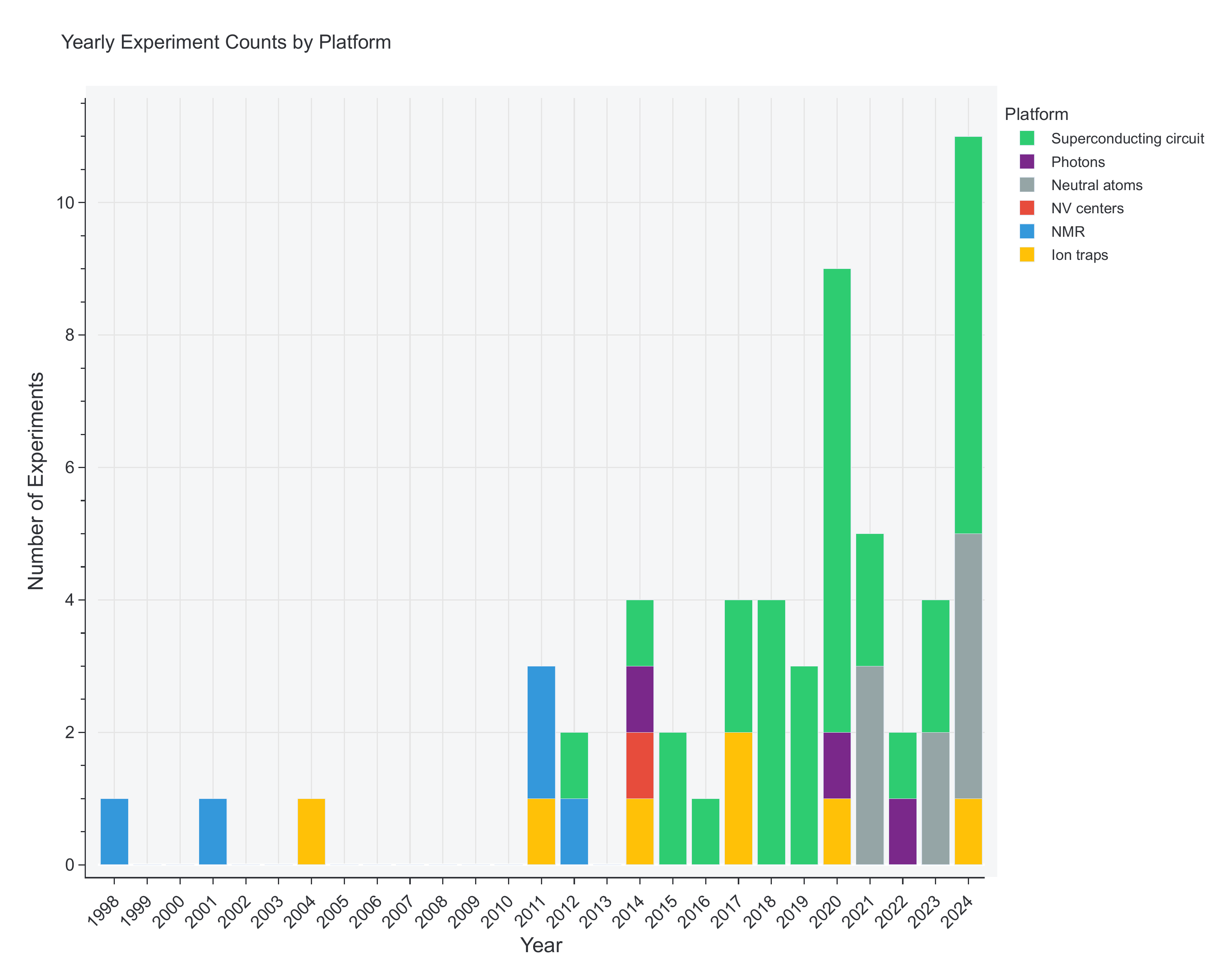}}

}

\caption{\label{fig-exp-yearly}\textbf{Yearly experiment counts by
platform.} Number of published QEC-related experiments per year,
categorized by platform.}

\end{figure}%

Complementing the cumulative view in Figure \ref{fig-exp-cumul}, Figure
\ref{fig-exp-yearly} displays the number of documented QEC-related
experiments on a year-by-year basis, categorized by the experimental
platform used. This non-cumulative representation highlights the annual
research output and reveals trends in platform activity over time, such
as the peak and subsequent decline of NMR contributions, the sustained
output from trapped ions, and the recent acceleration of research using
superconducting circuits and neutral atoms.

\subsubsection{Cumulative Growth of QEC Implementations by Code
Family}\label{cumulative-growth-of-qec-implementations-by-code-family}

\begin{figure}[ht]

\centering{

\pandocbounded{\includegraphics[keepaspectratio]{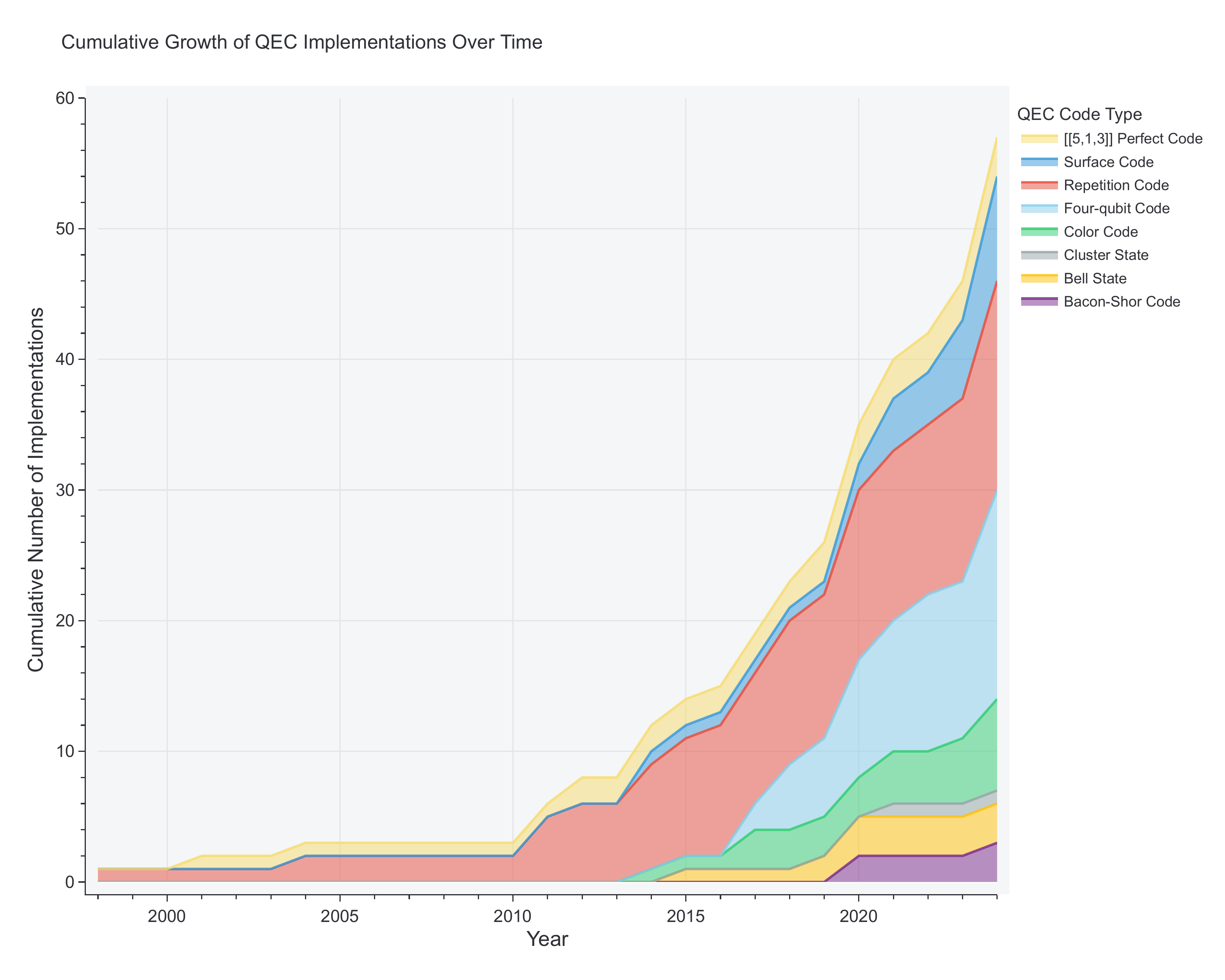}}

}

\caption{\label{fig-qec-growth}\textbf{Cumulative Growth of QEC
Implementations.} Total number of reported experimental implementations
for different QEC code families over time.}

\end{figure}%

Figure \ref{fig-qec-growth} illustrates the cumulative number of
reported experimental implementations specifically categorized by the
family of QEC code employed, tracked over time. This view highlights
which code families have received sustained attention. Early activity is
visible for central concepts demonstrated using Bell states and the
\([[5,1,3]]\) code. The plot shows significant recent growth in
implementations of surface codes and repetition codes (often used as
building blocks or testbeds for surface codes), along with substantial
effort dedicated to four-qubit codes.

\subsubsection{Timeline of Implementations}\label{sec-qec-timeline}

\begin{figure}[ht]

\centering{

\pandocbounded{\includegraphics[keepaspectratio]{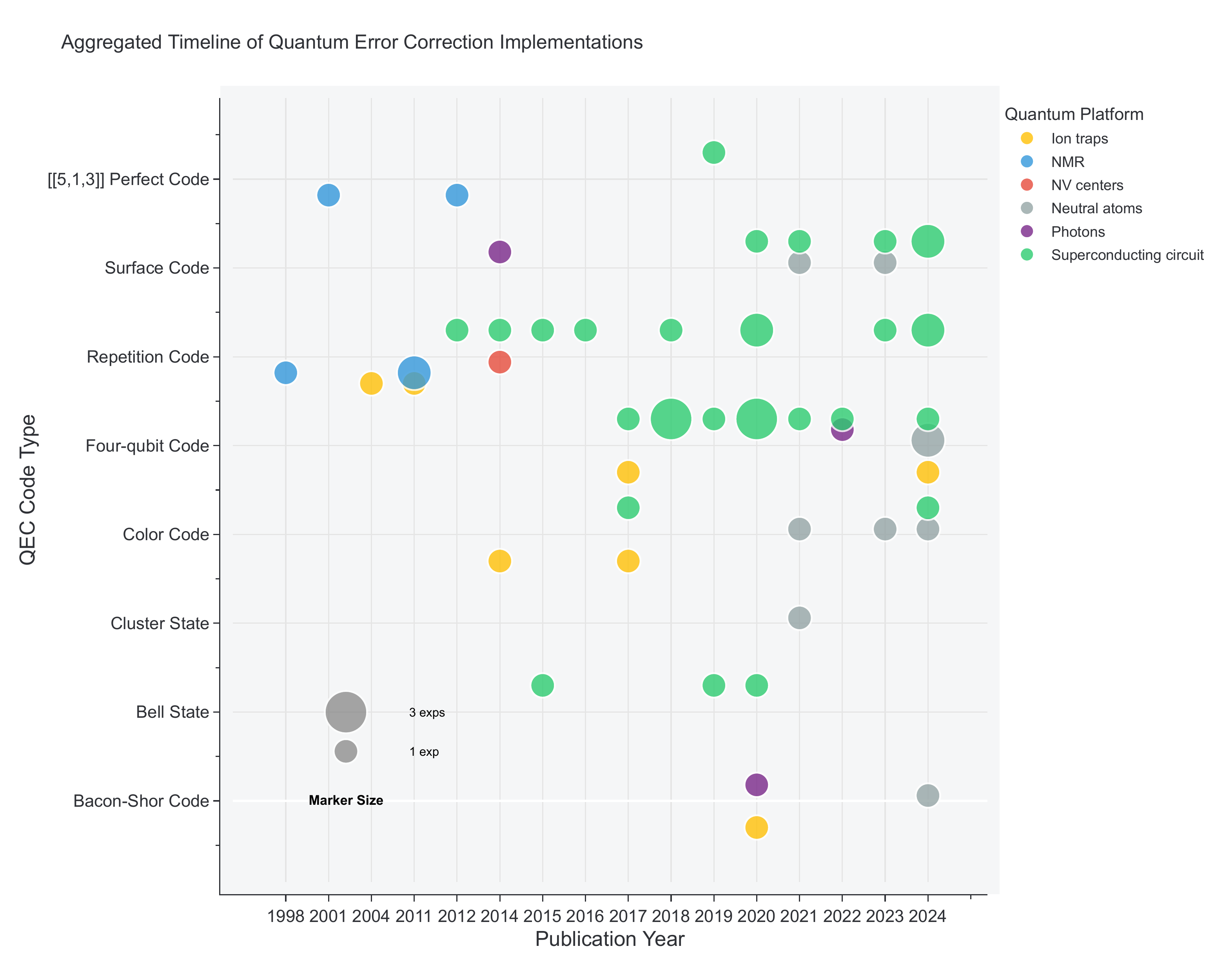}}

}

\caption{\label{fig-qec-timeline}\textbf{Timeline of QEC
Implementations.} Each point represents a reported experimental
implementation of a specific QEC code type (y-axis) on a particular
quantum platform (color/shape) in a given publication year (x-axis).}

\end{figure}

Figure \ref{fig-qec-timeline} provides a detailed chronological
perspective, mapping specific QEC code implementations (y-axis) to their
publication year (x-axis) and indicating the platform used via color and
shape. This visualization clearly shows the pioneering role of NMR in
early demonstrations, the consistent contributions from trapped ion
systems over many years, and the significant surge of activity in
superconducting circuits and neutral atoms, particularly in the last
decade. It also illustrates the progression towards implementing more
complex codes, such as surface and color codes, primarily on these
latter platforms.

\subsubsection{Distribution of Implementations Across Platforms and
Codes}\label{distribution-of-implementations-across-platforms-and-codes}

\begin{figure}[ht]

\centering{

\pandocbounded{\includegraphics[keepaspectratio]{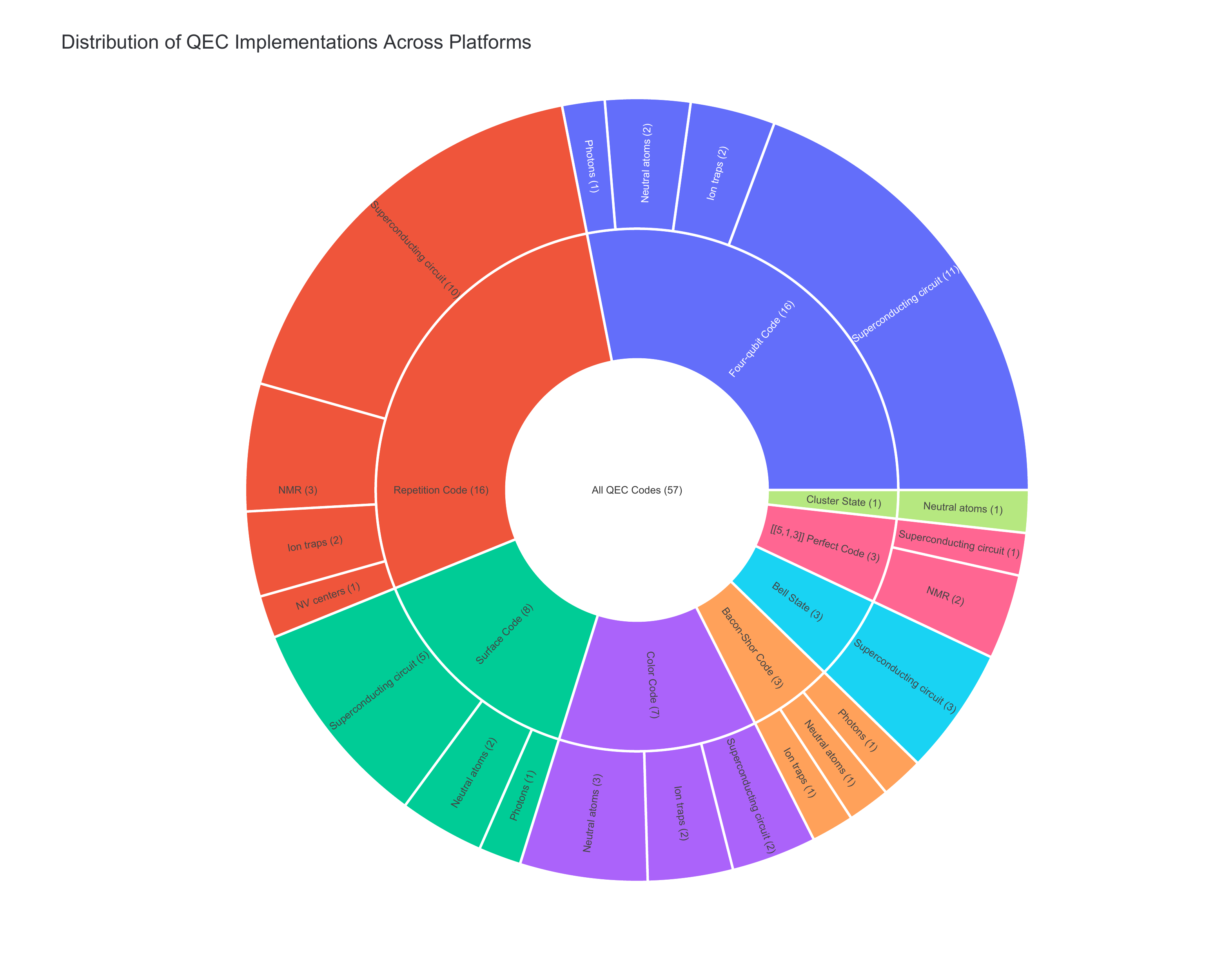}}

}

\caption{\label{fig-qec-sunburst}\textbf{Distribution of QEC
Implementations Across Platforms.} Sunburst chart showing the breakdown
of reported QEC code implementations (inner ring) by the physical
platform used (outer ring). Numbers indicate the count of publications
in the
\href{https://github.com/francois-marie/awesome-quantum-computing-experiments/blob/main/data/qec_exp.csv}{\texttt{qec\_exp}}
database.}

\end{figure}%

The distribution of QEC experimental efforts across different hardware
platforms and code types is summarized in the sunburst chart presented
in Figure \ref{fig-qec-sunburst}. The inner rings represent the physical
platforms, while the outer rings show the specific QEC codes
implemented, with segment sizes proportional to the number of documented
experiments (indicated by counts). This reveals that superconducting
circuits have hosted the largest volume and diversity of QEC experiments
to date, followed by trapped ions and, more recently, neutral atoms. It
also highlights that certain codes, notably repetition codes and
four-qubit codes, have been realized across multiple platforms, offering
opportunities for cross-platform comparison. In contrast,
implementations of larger-scale topological codes (e.g., surface, color
codes with higher distances) are currently concentrated on platforms
demonstrating higher available qubit counts.

\subsection{Goodness of Fit for Exponential Trends}\label{sec-fit}

This appendix details the methodology used to assess how well the
exponential trend lines, presented in Section~\ref{sec-phys-metrics} for
physical qubit benchmarks (Figure~\ref{fig-cohe},
Figure~\ref{fig-enterr}, Figure~\ref{fig-qcount}), fit the collected
historical experimental data.

\subsubsection{Methodology}\label{methodology}

To model the observed trends in coherence times (\(T_1\), \(T_2\)),
entanglement error, and physical qubit count over time (year, \(x\)),
we assumed an exponential relationship of the form: \(y = A \cdot B^x\)
where \(y\) represents the metric being analyzed. This exponential model
can be linearized by applying a (base-10) logarithm:
\(\log_{10}(y) = \log_{10}(A) + x \cdot \log_{10}(B)\) This
transformation yields a linear equation \(Y = c + m \cdot x\), where
\(Y = \log_{10}(y)\), the intercept \(c = \log_{10}(A)\), and the slope
\(m = \log_{10}(B)\).

Standard linear regression was performed on this transformed data (year
vs.~\(\log_{10}(\text{metric})\)) using the
\texttt{scipy.stats.linregress} function from the SciPy library
\cite{virtanenSciPy10Fundamental2020}. This function calculates the
best-fit line parameters (slope \(m\) and intercept \(c\)) and also
returns the Pearson correlation coefficient, denoted as \(r\).

The goodness of fit for this linear regression in log-space is
quantified by the coefficient of determination, \(R^2\), which is simply
the square of the Pearson correlation coefficient (\(R^2 = r^2\)). The
\(R^2\) value represents the proportion of the variance in the dependent
variable (\(\log_{10}(\text{metric})\)) that is predictable from the
independent variable (year). An \(R^2\) value ranges from 0 to 1, where
1 indicates that the linear model perfectly explains the variance in the
transformed data (corresponding to a perfect exponential fit for the
original data), and 0 indicates no linear correlation. Values close to 1
suggest that the exponential model provides a good description of the
observed historical trend.

\subsubsection{Results}\label{results}

The calculated \(R^2\) values obtained from the linear regression fits
performed for Figure~\ref{fig-cohe}, Figure~\ref{fig-enterr},
Figure~\ref{fig-qcount}, are summarized in Table~\ref{tbl-fit} below.
Fits were only calculated where sufficient data points (at least three)
were available for a given platform and metric.

\begin{longtable}[]{@{}
  >{\raggedright\arraybackslash}p{(\linewidth - 8\tabcolsep) * \real{0.2778}}
  >{\raggedright\arraybackslash}p{(\linewidth - 8\tabcolsep) * \real{0.3333}}
  >{\raggedright\arraybackslash}p{(\linewidth - 8\tabcolsep) * \real{0.0694}}
  >{\raggedright\arraybackslash}p{(\linewidth - 8\tabcolsep) * \real{0.1944}}
  >{\raggedright\arraybackslash}p{(\linewidth - 8\tabcolsep) * \real{0.1250}}@{}}
\caption{\textbf{Goodness of Fit (\(R^2\)) for Exponential Trends}. Note
the lower fit quality for neutral atoms \(T_1\) corresponding to the
different atomic species (Rubidium, Cesium, and Strontium) and
superconducting circuits \(T_2\) due to the different implementations
(transmons vs cavities).}\label{tbl-fit}\tabularnewline
\toprule\noalign{}
\begin{minipage}[b]{\linewidth}\raggedright
Metric
\end{minipage} & \begin{minipage}[b]{\linewidth}\raggedright
Platform
\end{minipage} & \begin{minipage}[b]{\linewidth}\raggedright
R\(^2\)
\end{minipage} & \begin{minipage}[b]{\linewidth}\raggedright
Time Scale
\end{minipage} & \begin{minipage}[b]{\linewidth}\raggedright
Std Error
\end{minipage} \\
\midrule\noalign{}
\endfirsthead
\toprule\noalign{}
\begin{minipage}[b]{\linewidth}\raggedright
Metric
\end{minipage} & \begin{minipage}[b]{\linewidth}\raggedright
Platform
\end{minipage} & \begin{minipage}[b]{\linewidth}\raggedright
R\(^2\)
\end{minipage} & \begin{minipage}[b]{\linewidth}\raggedright
Time Scale
\end{minipage} & \begin{minipage}[b]{\linewidth}\raggedright
Std Error
\end{minipage} \\
\midrule\noalign{}
\endhead
\bottomrule\noalign{}
\endlastfoot
Coherence Time (\(T_1\)) & Semiconductor & 0.954 & ×2 every 0.86y &
0.077 \\
& Superconducting circuit & 0.809 & ×2 every 0.99y & 0.032 \\
& Neutral atoms & 0.122 & ×2 every 5.89y & 0.079 \\
Coherence Time (\(T_2\)) & Neutral atoms & 0.880 & ×2 every 2.38y &
0.033 \\
& Ion traps & 0.860 & ×2 every 1.92y & 0.045 \\
& Superconducting circuit & 0.418 & ×2 every 1.94y & 0.045 \\
Entanglement Error & Semiconductor spins & 0.981 & ÷2 every 1.16y &
0.036 \\
& Ion traps & 0.952 & ÷2 every 2.40y & 0.011 \\
& Superconducting circuits & 0.933 & ÷2 every 2.56y & 0.016 \\
& Neutral atoms & 0.889 & ÷2 every 2.30y & 0.015 \\
Physical Qubit Count & Semiconductor spins & 0.983 & ×2 every 2.62y &
0.015 \\
& Ion traps & 0.955 & ×2 every 5.03y & 0.007 \\
& Neutral atoms & 0.948 & ×2 every 1.37y & 0.019 \\
& Superconducting circuit & 0.877 & ×2 every 2.58y & 0.025 \\
\end{longtable}

\subsubsection{Discussion}\label{discussion}

The \(R^2\) values presented in Table~\ref{tbl-fit} are generally high
(many above 0.85), suggesting that the exponential model captures a
significant portion of the variance in the historical trends for most
platforms and metrics analyzed. This supports the visual impression from
Figure~\ref{fig-cohe}, Figure~\ref{fig-enterr}, Figure~\ref{fig-qcount}
and provides quantitative justification for using exponential fits to
characterize the rates of progress (doubling/halving times).

However, certain caveats should be noted:

\begin{itemize}
\tightlist
\item
  The \(R^2\) value for Superconducting Circuit \(T_2\) time (0.418) is
  notably lower than others, reflecting that we included different
  families of implementations within the ``superconducting qubits''
  category (transmons vs cavities), and an outlier with particularly
  small \(T_2\) breaks the exponential fit for this platform.
\item
  Similarly, the lower \(R^2\) value for Neutral atoms \(T_1\) (0.122)
  reflects the inclusion of different atomic species (Rubidium, Cesium,
  and Strontium) with inherently different characteristics.
\item
  A high \(R^2\) value confirms that the exponential model fits the
  historical data well but does not guarantee that the trend will
  continue into the future, as discussed in Section~\ref{sec-us}
  regarding the limitations of extrapolation.
\end{itemize}

Overall, the analysis of \(R^2\) values reinforces the observation of
strong historical exponential trends in key physical qubit metrics
across major platforms, while also highlighting specific cases where the
model fit is weaker or based on limited data.

\end{document}